\documentclass[aip,apl,showpacs,showkeys,superscriptaddress,reprint]{revtex4-1}

\usepackage[utf8]{inputenc}
\usepackage[english]{babel}

\usepackage{amsmath,graphicx,siunitx,xcolor,textcomp}
\usepackage[caption=false]{subfig}
\usepackage[hidelinks]{hyperref}
\usepackage[english]{cleveref}

\newcommand{\inc}{\textup{inc}}
\newcommand{\crit}{\textup{crit}}
\newcommand{\spec}{\textup{spec}}
\newcommand{\imp}{\textup{imp}}
\newcommand{\out}{\textup{out}}

\newcommand{\bin}{\textup{bin}}
\newcommand{\beam}{\textup{beam}}
\newcommand{\submin}{\textup{min}}
\newcommand{\F}{\textup{F}}
\newcommand{\wg}{\textup{wg}}

\begin{document}
\title{X-ray scattering from whispering gallery mirrors}
\author{Nikolay \surname{Ivanov}}
\email[E-mail: ]{nd.ivanov@physics.msu.ru}
\affiliation{Department of Physics, Lomonosov Moscow State University, Moscow, 119991, Russia}
\author{Andrey \surname{Konovko}}
\email[E-mail: ]{konovko@physics.msu.ru}
\affiliation{Department of Physics, Lomonosov Moscow State University, Moscow, 119991, Russia}
\affiliation{International Laser Center, Lomonosov Moscow State University, Moscow, 119991, Russia}
\author{Anatoli \surname{Andreev}}
\email[E-mail: ]{av\_andreev@phys.msu.ru}
\affiliation{Department of Physics, Lomonosov Moscow State University, Moscow, 119991, Russia}

\date{\today}

\begin{abstract}
In this paper concave surface scanning method based on x-ray scattering in whispering gallery effect is considered. The capabilities of this method are studied based on ray-tracing computer simulation. The dependence of the output x-ray beam on the RMS roughness and minimum detectable imperfection height estimations were obtained.
\end{abstract}

\maketitle

\section{Introduction}
In third generation synchrotron light sources and in free-electron lasers, ultra-smooth mirrors with complex geometry are necessary to focus and control the X-ray beam \cite{tl10, ym12}. Also there is a great demand in modern astronomy for large ultra-smooth telescope mirrors with a diameter of 10 to 100 meters \cite{vm11,dw14}. In order to produce ultra-smooth mirrors with polishing techniques surface quality monitoring methods are required.

Currently, there is a number of surface scanning tools: stylus and optical profilometry \cite{ys11, cg10, hs15}, atomic force and tunnel microscopy \cite{mg15}, methods based on neutron, light and x-ray scattering \cite{ys15, zh15} and various interference schemes \cite{ng17}. They have different applications and give surface roughness information in different spatial frequency ranges.
For instance, optical profilometry scanning methods can be applied to mirrors with complex geometry, but the minimum longitudinal surface roughness size that can be detected is constrained in order of magnitude by the wavelength of the probing radiation \cite{yo15}. In mechanical profilometry stylus contacting the surface damages it to a certain extent \cite{ma07}.

In this paper the capabilities of the concave surface scanning method based on x-ray scattering in whispering gallery effect are studied. In the considered surface scanning method it’s proposed to investigate concave mirrors with x-ray beam, which “slides” over the surface due to whispering-gallery phenomenon. This phenomenon comes from acoustics where it's name originates. Whispering gallery is a type of effective x-ray light propagation over a concave mirror, where an x-ray beam reflects multiple times over the mirror with small grazing angle $\theta_{\inc}$ smaller than the critical grazing angle $\theta_{\crit} = |1-\varepsilon|^{1/2}$. By each act of beam reflection some of the light can be scattered from surface roughness. Scattered light can be absorbed by the mirror or propagate out of the surface at angles distinct from specular reflected light. Thus, light scattering leads to reduced efficiency of beam transfer and altered angular distribution of the output beam. This change in angular distribution contains information about the surface roughness. 

The transfer efficiency for a concave mirror with rotation angle $\psi$ (\cref{fig1}) and x-ray beam falling on the mirror with grazing angle $\theta_{\inc}$ equals\cite{kv95}:
\begin{equation}
    R_{\wg}(\theta_{\inc},\lambda) = \left(R_{\F}(\theta_{\inc},\lambda)\right)^N,\quad N = \left[\frac{\psi}{2\theta_{\inc}}\right]+1
\end{equation}
where $R_{\F}$ --- Fresnel reflectance, N --- number of ray reflections in whispering gallery propagation.

The transfer efficiency $R_{\wg}$ depends only on the mirror's rotation angle $\psi$ and by orders of magnitude higher compared to the single reflection case for $\theta_{\inc} > \theta_{\crit}$. This fact allows to apply considered scanning method to the concave mirrors with various sizes and shapes as well as to mirrors made of different materials. Moreover, the probing beam in whispering gallery propagation mode reflects from the mirror multiple times along it's trace. Thus, x-ray beam scans the entire surface in a single pass.

To present, the possibility of obtaining statistical roughness characteristics for cylindrical whispering gallery mirrors based on x-ray scattered light analysis has been investigated \cite{ab97}, and also the problem of x-ray beam scattering on spherical mirror without taking scattering from surface roughness into consideration has been studied \cite{yr08}.

Two practical problems are posed in this paper:
\begin{itemize}
	\item Obtaining the statistical characteristics of the surface roughness from the angular distribution of the output beam.
    \item Surface defects detection.
\end{itemize}
All these problems are studied based on ray tracing computer simulation \cite{hh89}. The dependence of the output x-ray beam on the statistical surface roughness characteristics and minimum imperfection size estimations that can be detected using the x-ray surface investigation method considered in this paper were obtained.
\section{Computer simulation of X-ray beam scattering}
\subsection{Experimental setups}
\begin{figure}[!t]
	\subfloat[\label{fig1:1}]{\includegraphics[width=0.48\linewidth]{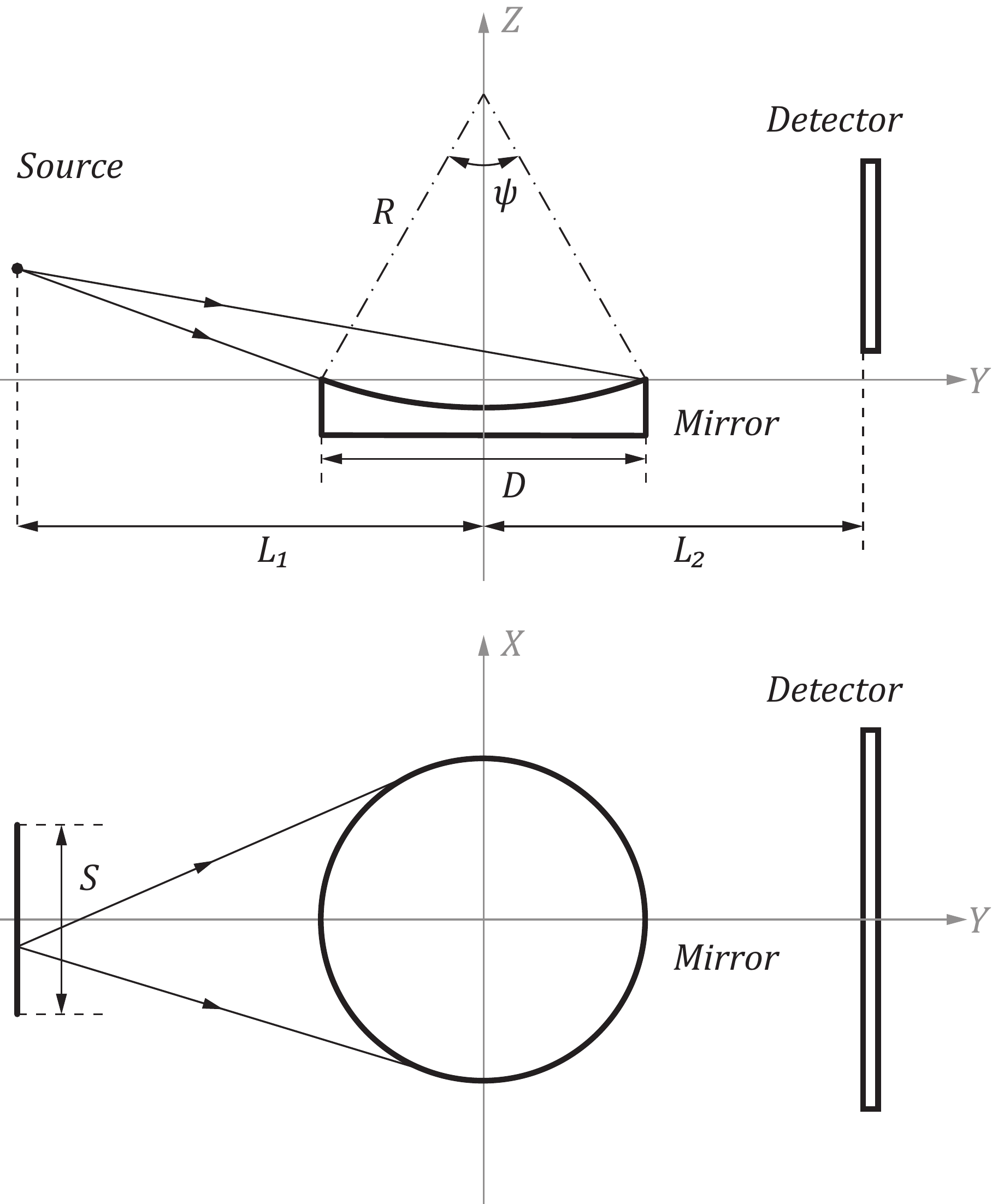}}\hfill
    \subfloat[\label{fig1:2}]{\includegraphics[width=0.48\linewidth]{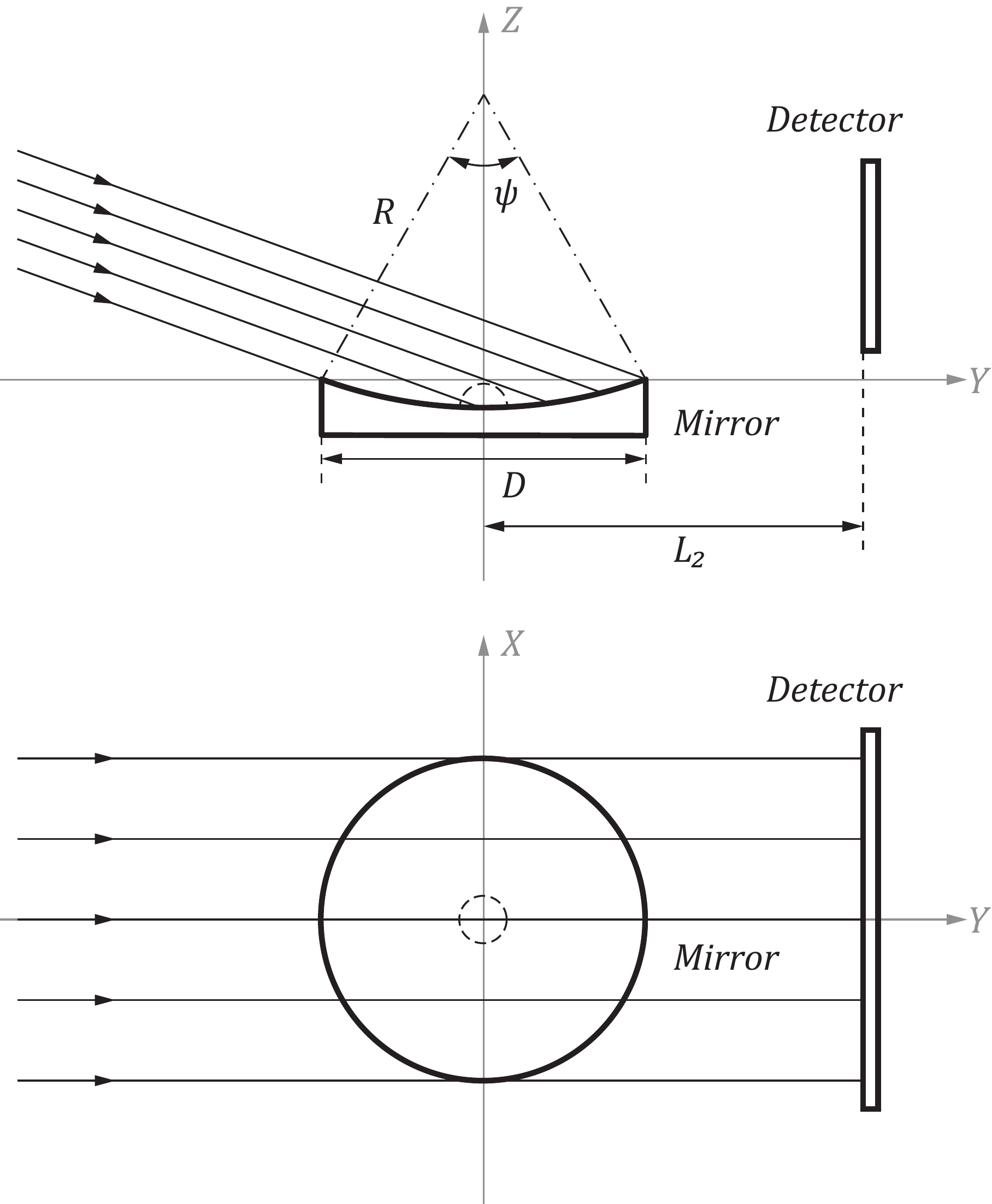}}
    \caption{Experimental setup for concave spherical mirrors study; All ray tracing simulations were performed with the following parameter values: $L_1 = \SI{195}{\milli\metre}$, $L_2 = \SI{185}{\milli\metre}$, $S = \SI{60}{\milli\metre}$}
    \label{fig1}
\end{figure}

Let us consider two experimental setups for concave mirrors study (\cref{fig1}). In the former (\cref{fig1:1}) the non-collimated beam from extended source located at a tangent to a mirror falls on the concave mirror, and in the latter (\cref{fig1:2}) a collimated beam falls at a tangent on a concave mirror. The incident x-ray light was considered monochromatic with Copper $K_\alpha$ wavelength.

We note that the latter scheme is more informative but requires careful alignment of the incident beam with respect to the surface. The latter scheme also allows one to reconstruct the shape of a surface defect on the mirror from tilt-series of the x-ray beam, which was done by \citeauthor{yr08} for a fingerprint on the spherical mirror \cite{yr08,yb12}. The surface defects detection problem was posed to examine the sensitivity of the given technique.

Surface defects were modeled by truncated spheres placed on a mirror(\cref{fig1:2}). They were specified by surface coordinates ${x_\textup{c}, y_\textup{c}}$,  the width $d_{\imp}$ and the height $h_{\imp}$ of the defect.
\subsection{Ray tracing algorithm}
To establish whether it is possible to investigate concave mirrors by the output beam measurements, a computer simulation of an x-ray beam scattering on a concave surface based on the ray tracing algorithm was carried out. In ray tracing technique the beam is treated as a bundle of unit power rays (in the order of $10^8\text{--}10^9$ in our calculations). Every ray is traced separately. With each act of ray incidence three events may happen:
\begin{enumerate}
	\item The ray reflects specularly from the surface. The probability of this event is equal to specular reflection coefficient $R_{\spec}$.
    \item The ray scatters on the surface roughness. The probability is equal to total integrated scatter $TIS$. In this case a pair of random numbers, which are the scattering angles $\theta$ and $\varphi$, are generated. The former is grazing angle in incidence plane and the latter is rotation angle in the azimuthal plane. The distribution density is defined by the normalized scattering indicatrix $\Phi(\theta,\varphi;\theta_{\inc})/TIS$.
    \item The ray may be absorbed with the probability $1 - R_{\spec} - TIS$. In this case the computer begins to trace a new ray.
\end{enumerate}

To choose one of three possible outcomes a random number generator in the interval $[0;1]$ is used. A number falling into the interval $[0; TIS]$, or $[TIS; TIS + R_{\spec}]$, or $[TIS + R_{\spec}; 1]$ determines which event is realized, respectively. The procedure repeats until the ray escapes from the opposite side of the mirror.

\subsection{Surface roughness modulation}

In our calculations the specular reflection coefficient $R_{\spec}$ and the total integrated scatter $TIS$ in Kirchhoff approximation were used, in which they are defined by Fresnel reflectance $R_{\F}(\theta_{\inc},\lambda)$ and Debye–Waller factor \cite{ma07, kv95}.

The statistical properties of surface roughness are described by the power spectral density (PSD) function, which is the Fourier transform of the autocorrelation function. We used ABC-model PSD-function characterized by root-mean-square (RMS) roughness $\sigma$, correlation length $\xi$ and fractal dimension of a surface $D$ \cite{bd10}:
\begin{equation}
    PSD^\textup{2D}_\textup{ABC}(\pmb{\nu}) = \frac{\sigma^2\xi^2\alpha}{\pi(1 + \pmb{\nu}^2\xi^2)^{1+\alpha}}
\end{equation}
where $\alpha$ defines the fractal dimension $D_{\textup{frac}} = 3 - \alpha$, $0 < \alpha < 1$.

In the first-order perturbation theory the scattering indicatrix $\Phi(\theta,\varphi;\theta_{\inc})$ is defined by PSD-function \cite{kv95}.

\section{Discussion of the computer simulation results}
\subsection{Influence of surface roughness scattering on spatial distribution of the output beam}
\begin{figure}
    \subfloat[$\sigma = \SI{0}{\nano\metre}$  \label{fig2:1}]{\includegraphics[height=3.8cm]{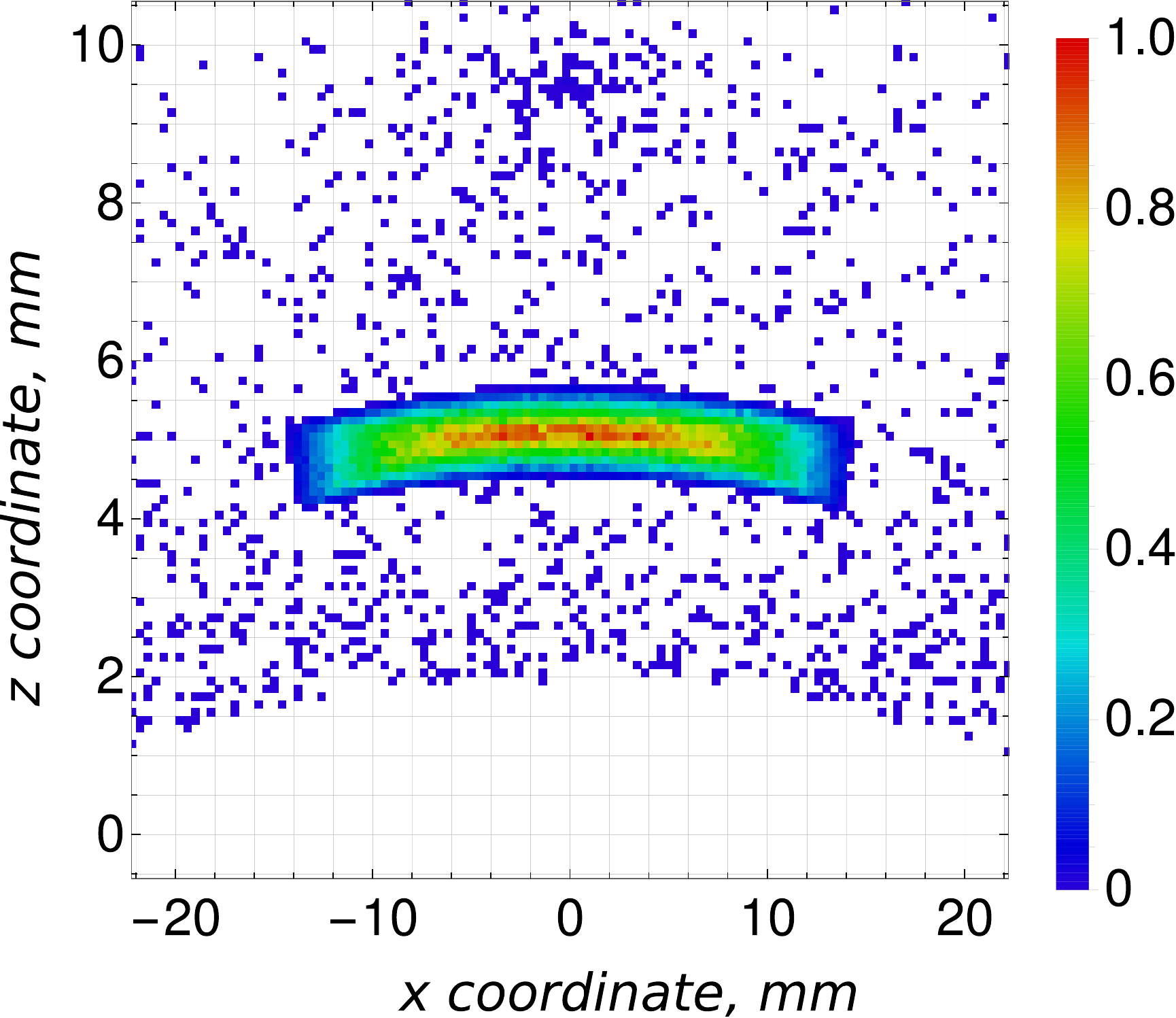}}\hfill
    \subfloat[$\sigma = \SI{1.2}{\nano\metre}$\label{fig2:2}]{\includegraphics[height=3.8cm]{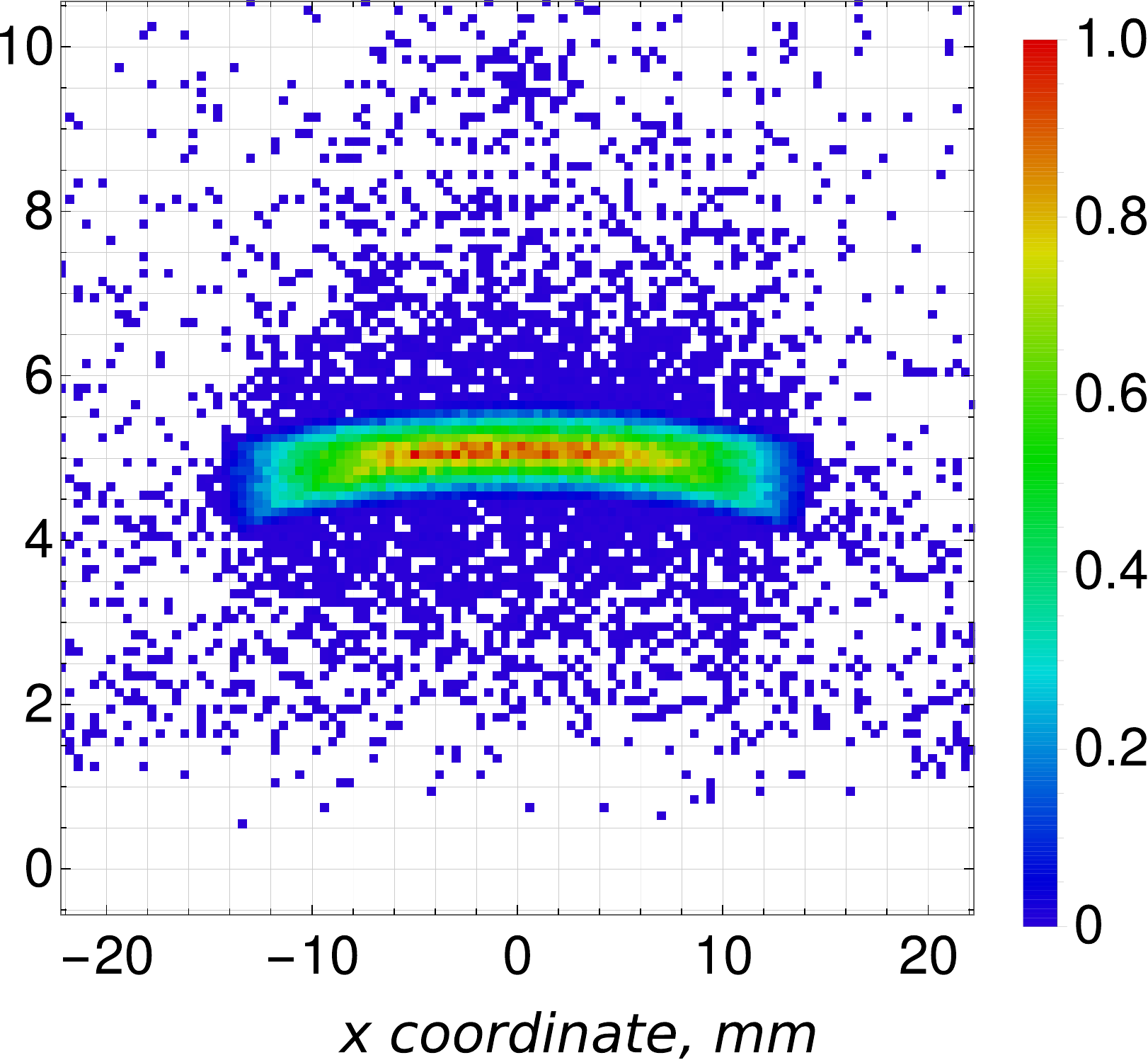}}
    \vfill
	\subfloat[$\sigma = \SI{0}{\nano\metre}$  \label{fig2:3}]{\includegraphics[height=3.8cm]{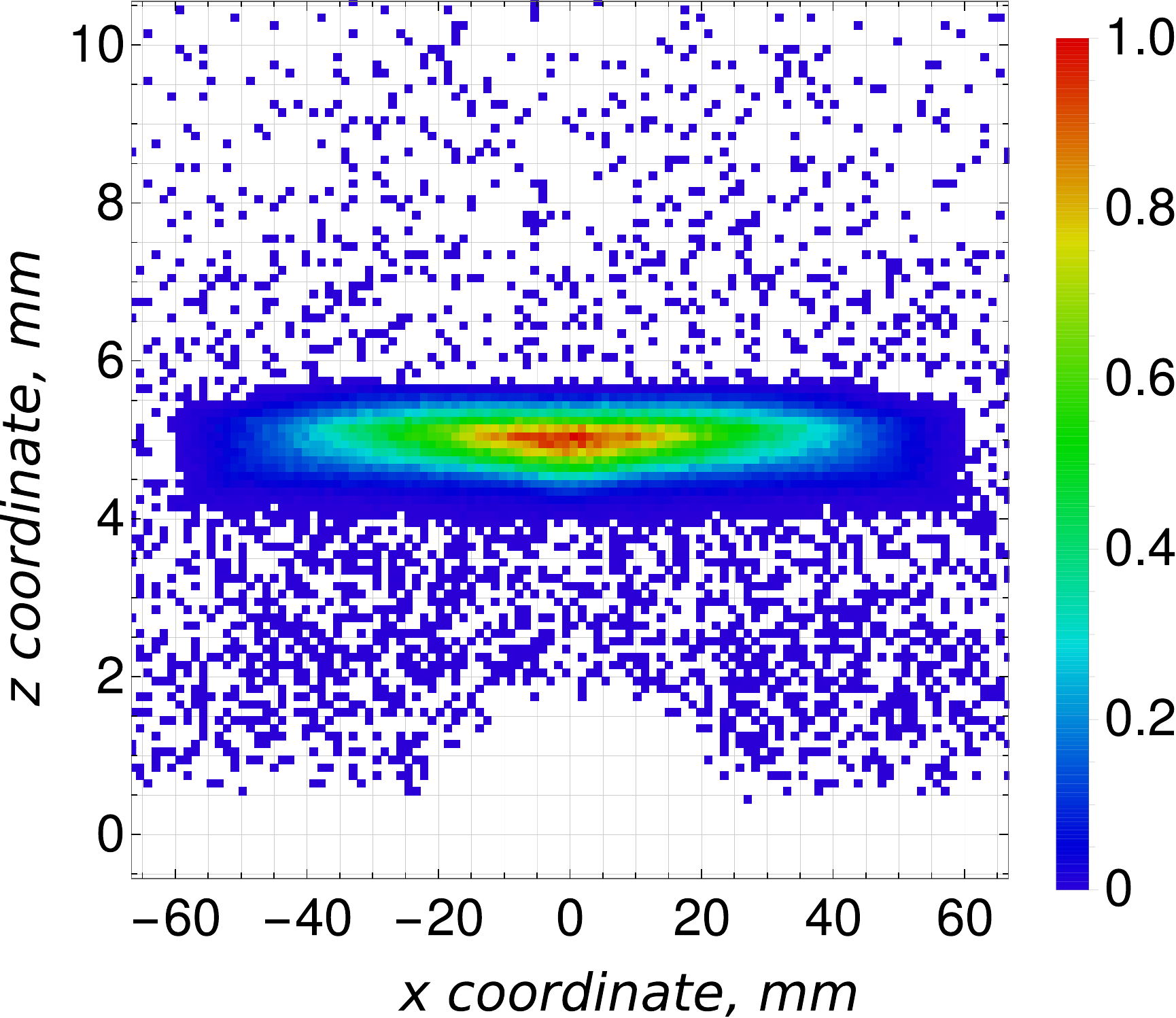}}\hfill
    \subfloat[$\sigma = \SI{1.2}{\nano\metre}$\label{fig2:4}]{\includegraphics[height=3.8cm]{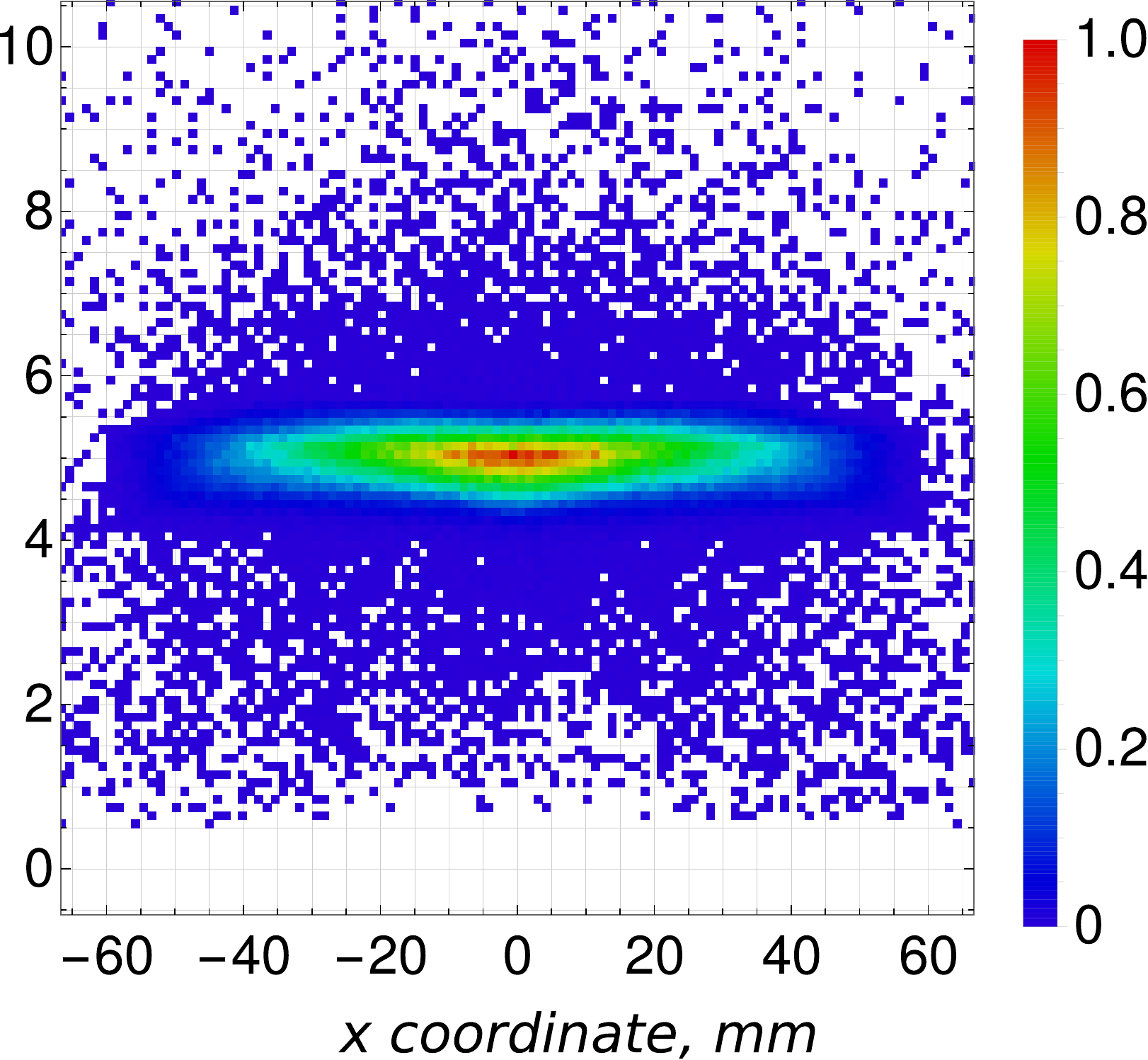}}
    \caption{The intensity distribution in the detector plane of the output beam scattered on a spherical mirror with different root-mean-square roughness $\sigma$ for cases of collimated (a, b) and non-collimated (c, d) incident beam. Following parameters were used: $R = \SI{1000}{\milli\metre}$, $D = \SI{60}{\milli\metre}$ and $\psi = \ang{3.44}$}
    \label{fig2}
\end{figure}
The \cref{fig2} demonstrates results of x-ray beam scattering on a ideally smooth spherical mirror and on a rough spherical mirror in cases of collimated (\cref{fig2:1,fig2:2}) and non-collimated incident beam (\cref{fig2:3,fig2:4}). In these figures we see bright spot of x-ray light propagated over the mirror in whispering gallery regime and background signal consisted of x-ray light scattered on the surface roughness and x-ray light reflected from the mirror only one time with grazing angle $\theta_{\inc} > \theta_{\crit}$. In case of non-collimated incident beam the output beam is highly divergent with divergence angle $\theta_\textup{div} = \frac{S+D}{2 L_1}=\ang{17.63}$.

\begin{figure}[t]
	\includegraphics[width=\linewidth]{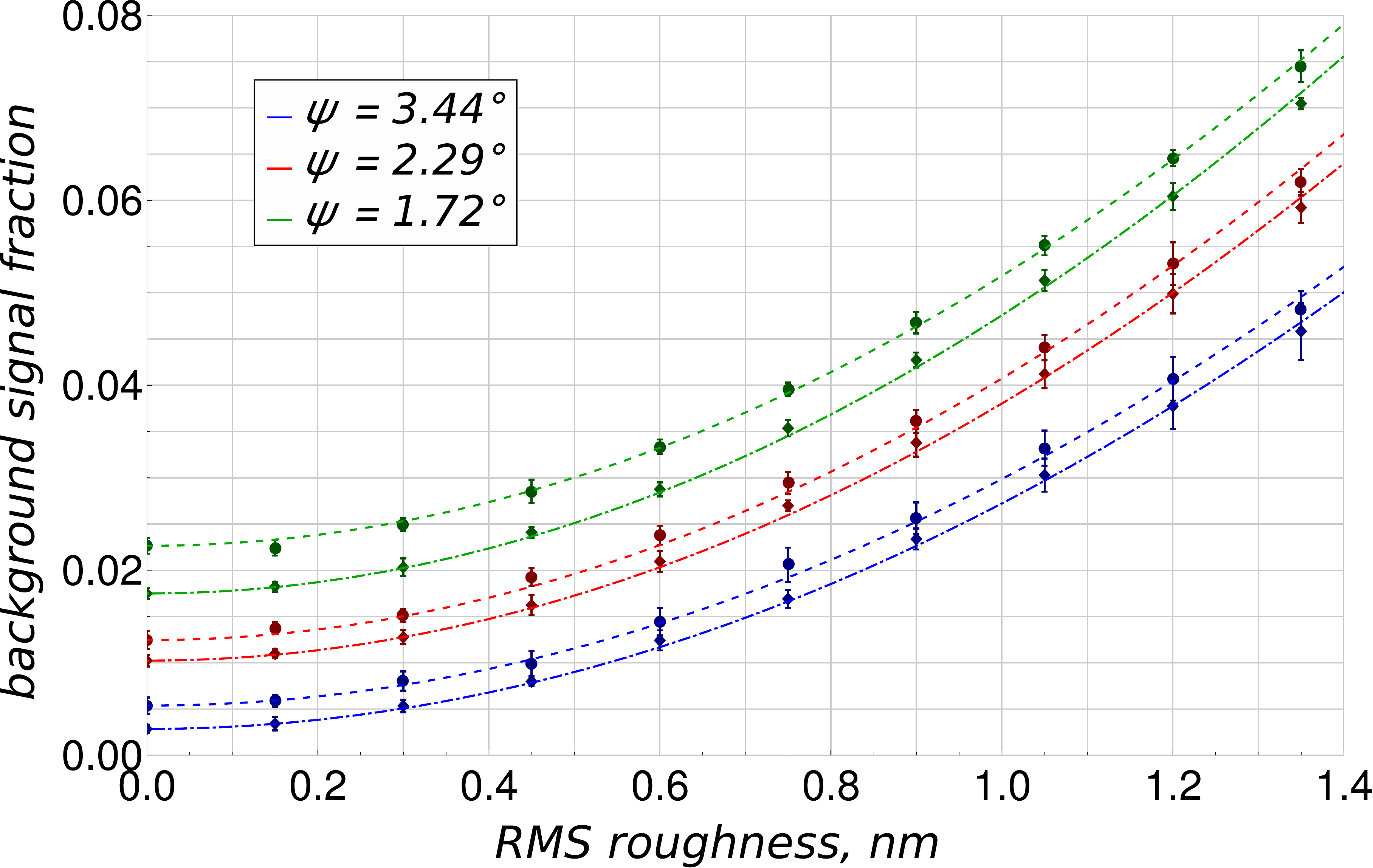}
    \caption{Background signal fraction dependence on the RMS roughness: interpolations (lines) and ray tracing simulation results (points) in cases of collimated (round points and dashed lines) and non-collimated incident beam (rectangular point and dot-and-dashed lines) for mirrors with different geometries: green --- $R = \SI{1000}{\milli\metre}$, $D = \SI{60}{\milli\metre}$; red --- $R = \SI{1500}{\milli\metre}$, $D = \SI{60}{\milli\metre}$; blue --- $R = \SI{2000}{\milli\metre}$, $D = \SI{60}{\milli\metre}$}
    \label{fig3}
\end{figure}

\begin{figure}[t]
	\includegraphics[width=\linewidth]{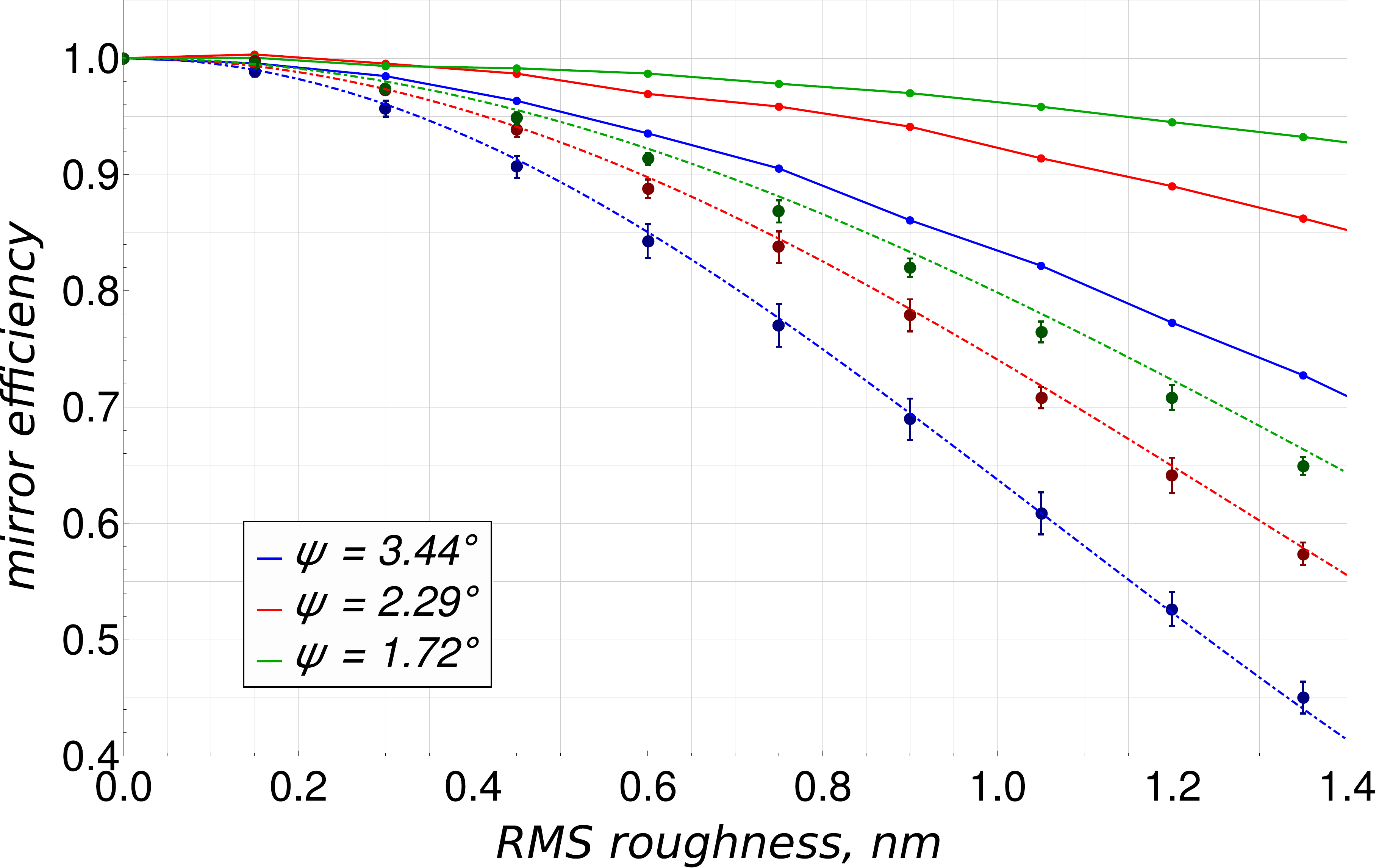}
    \caption{Concave spherical mirror efficiencies for total light (solid line) and specularly reflected light (dashed line); Mirrors are the same as in \cref{fig3}}
    \label{fig4}
\end{figure}

The \cref{fig3} demonstrates the dependences of the background signal fraction on the RMS roughness. These dependences were obtained from numerical simulation results of x-ray beam scattering on spherical mirrors with different geometries and different RMS roughness in cases of collimated and non-collimated incident beam. The background signal fraction was defined as the ratio of the background signal power to the total output beam power. These graphs correspond to the dependence of the total integrated scatter $TIS$ on the RMS roughness with a high accuracy, and the dependences of the background signal fraction for different mirrors and different incident beam geometries differ by a constant. This constant is the background power of one time reflected light.\\
Besides, a mirror absorbs the most part of scattered light, thus surface roughness scattering decreases efficiency of the mirror. Computer simulation has shown that the mirror efficiencies for the two considered incident beam configurations are equal to each other within the error margin. The \cref{fig4} depicts mirror efficiency dependences for total light and specularly reflected light in the case of collimated incident beam for different spherical mirrors. \citeauthor{kv95} derived estimations for losses of specularly reflected light in whispering gallery propagation mode due to scattering on surface roughness \cite{kv95}. These estimations correspond to computer simulation results (\cref{fig4}):
\begin{equation}
	R_{\spec}^{\out}(\theta_{\inc},\psi,\sigma)\approx\exp\left[-\left(\frac{4\pi\sin^2\theta_{\inc}}{\lambda}\right)^2\frac{\psi}{2\theta_{\inc}}\right]
\end{equation}
\subsection{Influence of surface imperfections on the output beam}
\begin{figure}[t]
	\subfloat[\label{fig5:1}]{\includegraphics[width=0.48\linewidth]{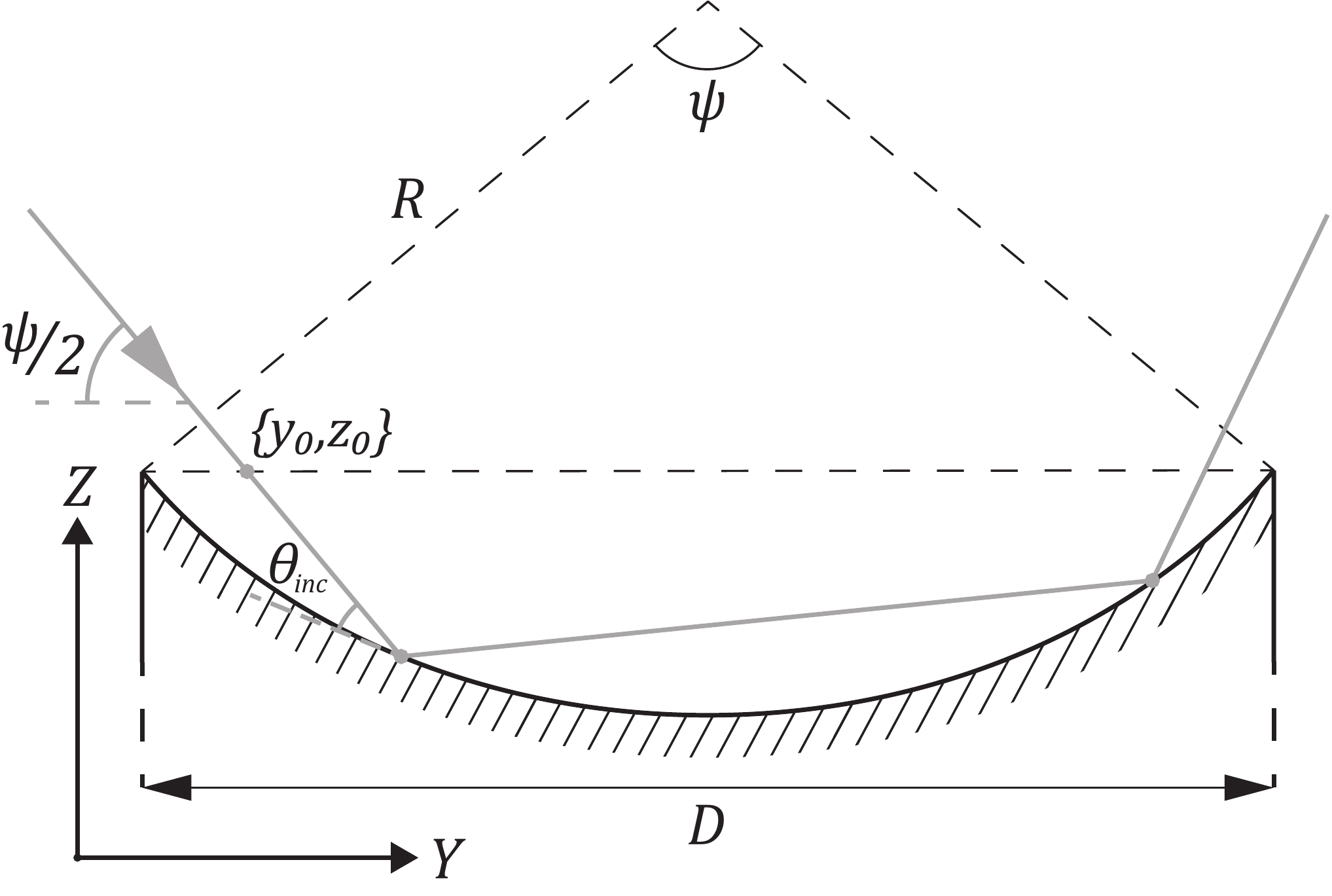}}
    \subfloat[\label{fig5:2}]{\includegraphics[width=0.48\linewidth]{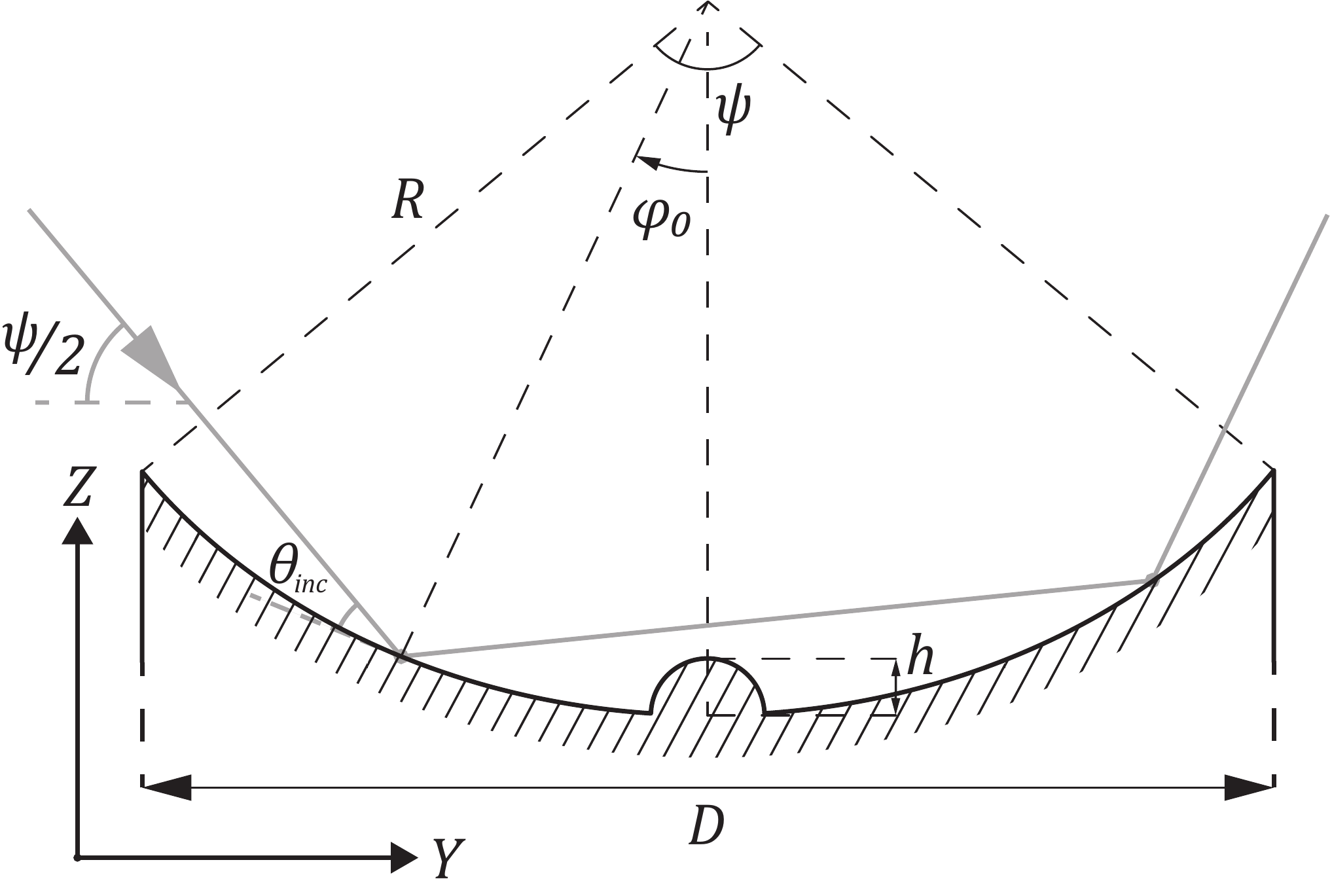}}
    \caption{Ray-tracing schemes for the cylindrical mirror without a surface imperfection (a) and in the presence of an imperfection (b)}
    \label{fig5}
\end{figure}

A defect placed on a concave mirror will absorb a part of x-ray light propagating along the mirror in whispering gallery mode. From the attenuation of whispering gallery bright spot in the output beam we can reconstruct the defect. The probability of the beam falling on the defect depends on the grazing angle at which the beam propagates along the mirror in the whispering gallery mode. To analytically find the output beam rays distribution from the grazing angle, we consider the case of a collimated beam falling on a ideally smooth cylindrical mirror (two-dimensional scattering problem) with radius of curvature $R$ and diameter $D$ (\cref{fig5:1}). In this configuration the grazing angle $\theta_{\inc}$ in the approximation of small rotation angle $\psi = 2\arcsin(D/2R)\approx D/R\ll 1$ equals:
\begin{equation}
	\centering
	\theta_{\inc} = \frac{\sqrt{(y_0 + D/2)D}\cos(\psi/2)}{R}
    \label{eq1}
\end{equation}

The incident beam is collimated, so $y_0$ is uniformly distributed along the interval $[-D/2,D/2]$. Then from the \eqref{eq1} the grazing angle distribution of output rays $f_{\out}(\theta_{\inc})$ could be derived:
\begin{equation}
	f_{\out}(\theta_{\inc})\cong
    \begin{cases}
    	C_{\wg}R_{\wg}(\theta_{\inc},\lambda)\theta_{\inc},	&	\theta_{\inc} < \theta_{\crit}\\
        0,									&	\theta_{\inc} > \theta_{\crit}
    \end{cases}
\end{equation}

where $C_{\wg} = (\int_0^{\theta_{\crit}}R_{\wg}(\theta_{\inc},\lambda)\theta_{\inc}\,d\theta_{\inc})^{-1}$ --- the normalizing constant.

\begin{figure}[t]
	\includegraphics[width=\linewidth]{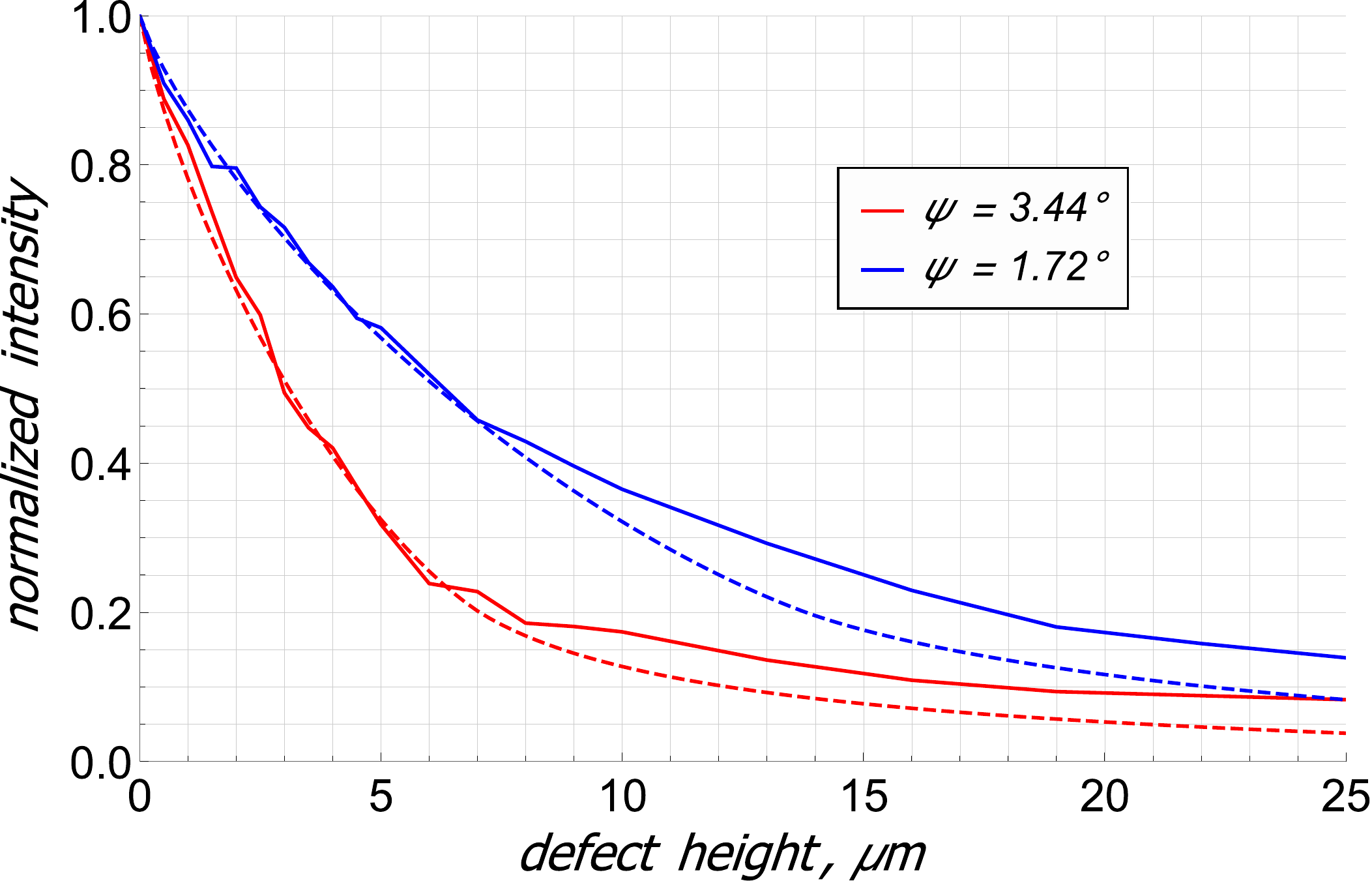}
    \caption{The output beam intensity for the mirrors with different rotation angles: the dotted line --- theoretical estimates (based on the \eqref{eq3}), solid --- computer simulation results; Mirrors geometries are the same as in \cref{fig3}}
	\label{fig6}
\end{figure}

Next, we consider a surface defect of height $h_{\imp}$ placed in the center of the mirror $\{0,0\}$ (\cref{fig5:2}). A ray propagating along the mirror in whispering gallery mode with a grazing angle $\theta_{\inc}$ is determined by the point of incidence on the mirror $\varphi_0\in[0,2\theta_{\inc}]$ before the defect. For the considered configuration, the probability of a ray passing past the imperfection $P(h_{\imp},\theta_{\inc})$ and the intensity of the output beam $J_{\out}(h_{\imp})$ normalized to the output intensity without a defect in the approximation $\theta_{\inc},\varphi_0\ll 1$ could be obtained (\cref{fig6}):
\begin{subequations}
\begin{equation}
	P(h_{\imp},\theta_{\inc}) = \frac{\sqrt{\theta_{\inc}^2-2h_{\imp}/R}}{\theta_{\inc}}
    \label{eq2}
\end{equation}
\begin{equation}
	J_{\out}(h_{\imp}) = \int_{\sqrt{\frac{2 h_{\imp}}{R}}}^{\theta_{\crit}}f_{\out}(\theta_{\inc})P(h_{\imp},\theta_{\inc})d\theta_{\inc}
    \label{eq3}
\end{equation}
\end{subequations}

The analytic output beam attenuation $J_{\out}(h_{\imp})$ was compared with ray tracing simulation of collimated beam falling on a rough spherical mirrors with defect placed in the center of the mirror (\cref{fig1:2}). The \cref{fig6} demonstrates the output beam intensity in surface imperfection localization area for imperfections with different heights $h_{\imp}$ and spherical mirrors with different rotation angles $\psi$. The output beam intensities at defect heights $h_{\imp}\gtrsim\SI{7}{\micro\metre}$ are close for different mirrors and do not correspond to theoretical estimations. At given defect heights, the output x-ray light in defect localization area corresponds mostly to one time reflected light, which is not considered in the foregoing analysis.

In order to estimate the minimum detectable defect height $h_{\submin}$, we use the detection limit \cite{mc80}:
\begin{equation}
    \langle I_{\out}\rangle - I_{\out}^{\imp} > 3\sigma_{I_{\out}}
\end{equation}

where $\langle I_{\out}\rangle$ --- the mean output intensity, $I_{\out}^{\imp}$ --- the output intensity in defect localization area and $\sigma_{I_{\out}}$ --- the standard deviation of the output intensity.

The output beam is registered by a matrix sensor, the pixel size $\Delta_{\bin}$ is chosen equal to the width of the defect $d_{\imp}$ to register the imperfection with minimal noise. In this case, the standard deviation, normalized to the mean output intensity, equals:
\begin{equation}
	\sigma_{I_{\out}}/\langle I_{\out}\rangle = \sqrt{\frac{\Delta_{\beam}/\Delta_{\bin} - 1}{N - \Delta_{\beam}/2 \Delta_{\bin} + 1}}
\end{equation}
where $\Delta_{\beam}$ --- the transverse width of the output beam in the case of the collimated incident beam:
\begin{multline}
	\Delta_{\beam}\approx\frac{\sqrt{D^2 - 4R^2\sin^2\alpha}}{\cos\alpha},\\
    \alpha = \arcsin\left(\frac{D}{2R}\right) - \frac{\theta_{\crit}}{2}
\end{multline}

\begin{figure}[t]
	\subfloat[\label{fig7:1}]{\includegraphics[width=0.48\linewidth]{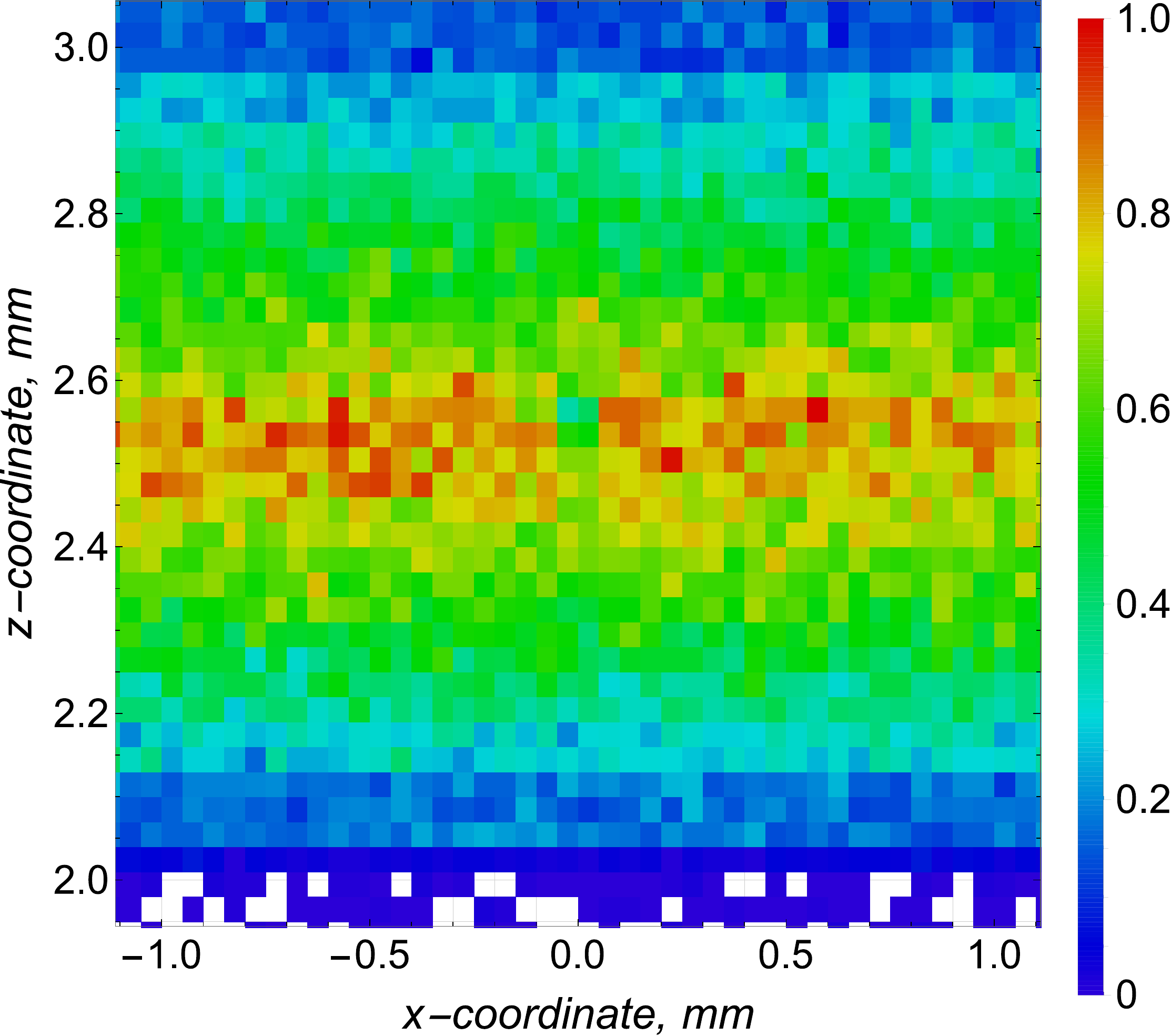}}\hfill
    \subfloat[\label{fig7:2}]{\includegraphics[width=0.48\linewidth]{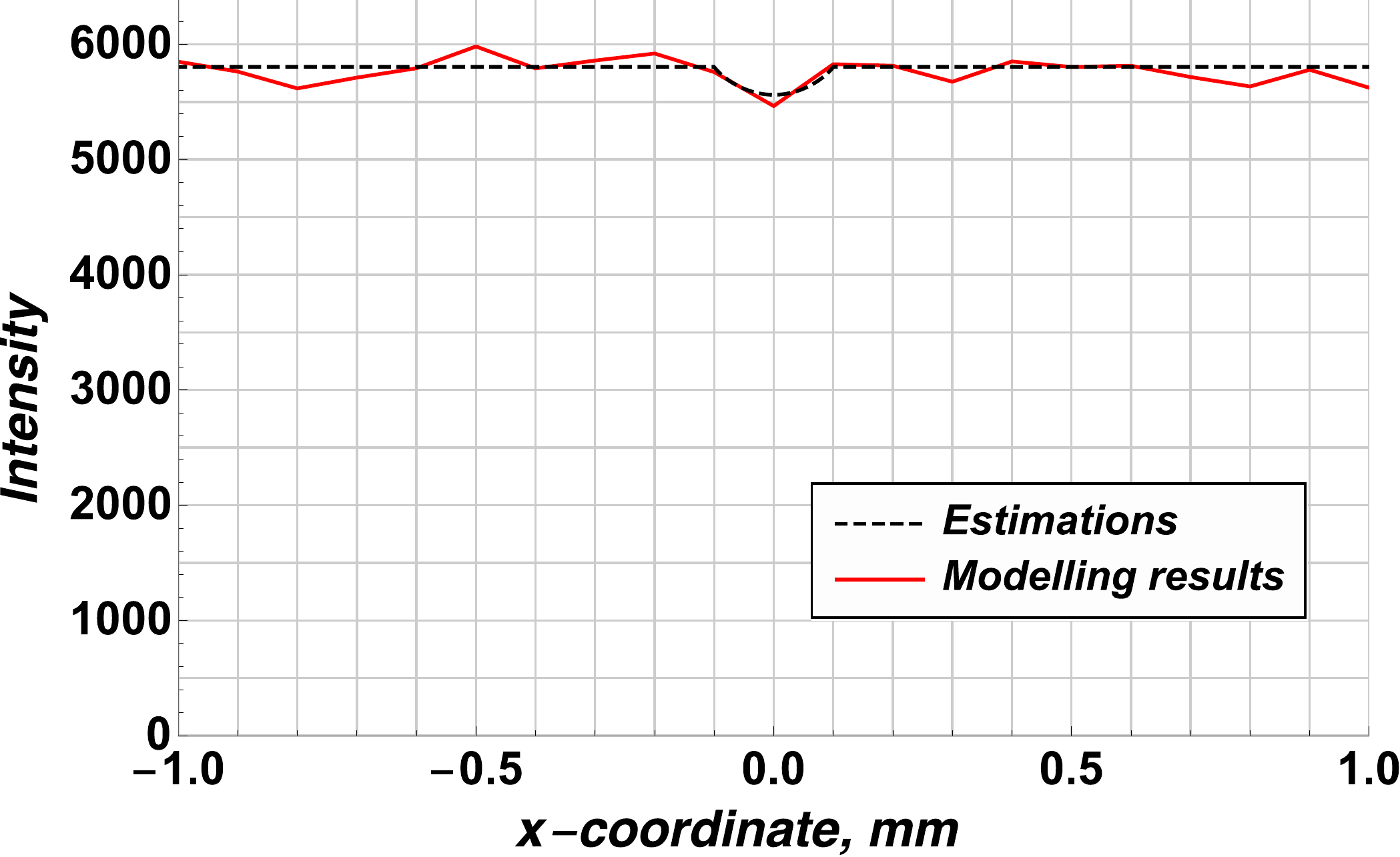}}
    \caption{The output beam intensity for the mirror and the defect with following parameters: the mirror radius of curvature $R = \SI{2}{\metre}$, the mirror diameter $D = \SI{60}{\milli\metre}$, the defect width $d_{\imp} = \SI{100}{\micro\metre}$, the defect height $h_{\imp} = \SI{344}{\nano\metre}$; (a) --- in the detector plane, (b) --- in the detector plane as a function of the transverse coordinate}
    \label{fig7}
\end{figure}

The detection condition allows us to estimate the minimum detectable defect height, for a mirror with a radius of curvature $R = \SI{2}{\metre}$ and diameter $D = \SI{60}{\milli\metre}$ and a defect of width $d_{\imp} = \SI{100}{\micro\metre}$, the height is $\SI{344}{\nano\metre}$. The output beam intensity  in the detector plane at the foregoing parameters is shown on \cref{fig7}. In the center of the \cref{fig7:1}, a region with a lower intensity is clearly visible, which corresponds to the defect localization area of width $d_{\imp}$. The latter figure (\cref{fig7:2}) depicts the output beam intensity histogram with bin width $\Delta_{\bin} = d_{\imp}$. The output intensity in the defect localization zone satisfies the detection limit $\langle I_{\out}\rangle - I_{\out}^{\imp} = 3.21\sigma_{I_{\out}} > 3\sigma_{I_{\out}}$.
\section{Conclusions}
In this paper, a ray-tracing algorithm with surface roughness scattering was outlined, a ray-tracing program for spherical mirrors was developed \footnote{\url{https://github.com/simply-nicky/raytrace}}, computer simulations of x-ray beam scattering on  spherical mirrors with and without a surface defect were performed. The dependences of background signal and mirror efficiency on the RMS roughness were obtained. For the case of a surface defect placed on the mirror, estimates of the minimum detectable defect height and the output beam intensity for a defect localization area were made. Thus, for a spherical mirror with a radius of curvature $R = \SI{2}{\metre}$ and diameter $D = \SI{60}{\milli\metre}$ and a defect of width $d_{\imp} = \SI{100}{\micro\metre}$, the minimum detectable defect height is $\SI{344}{\nano\metre}$. At a given defect height, the output beam intensity in the defect localization area satisfies the detection limit. The obtained analytic estimations are in good agreement with the results of numerical simulation.
\section{Acknowledgements}
This research was partially supported by the Russian Foundation for Basic Research grants \textnumero 18-02-00528 and \textnumero 16-29-11763. We thank colleagues from reflectometry and small-angle scattering laboratory of Crystallography Institute Russian Academy of Sciences and especially Dr. V. E. Asadchikov who provided insight and expertise in x-ray optics that greatly assisted the research. 
\bibliographystyle{aipnum4-1}
\bibliography{main}

\end{document}